\documentclass[11pt,a4paper]{article}
\usepackage{jheppub}

\usepackage{amsmath}
\usepackage{latexsym}

\title{\Large{QCD gauge symmetries through Faddeev-Jackiw symplectic method}}
\author[a,b]{Everton M. C. Abreu,}
\author[b]{Albert C. R. Mendes,}
\author[c]{Clifford Neves,}
\author[b]{Wilson Oliveira,}
\author[b]{Rodrigo C. N. Silva.}

\affiliation[a]{Grupo de F\'isica Te\'orica e Matem\'atica-F\' isica, Departamento de F\'{\i}sica, \\
Universidade Federal Rural do Rio de Janeiro\\
BR 465 km 07, 23890-971, Serop\'edica, RJ, Brazil}
\affiliation[b]{Departamento de F\'{\i}sica, ICE, Universidade Federal de Juiz de Fora,\\
36036-330, Juiz de Fora, MG, Brazil}
\affiliation[c]{Departamento de Matem\'atica e Computa\c{c}\~ao, Universidade do Estado do Rio de Janeiro\\
Rodovia Presidente Dutra, km 298, 27537-000, Resende, RJ, Brazil}

\emailAdd{evertonabreu@ufrrj.br}
\emailAdd{albert@fisica.ufjf.br}
\emailAdd{clifford@fat.uerj.br} 
\emailAdd{wilson@fisica.ufjf.br}
\emailAdd{rodrigocnsilva@fisica.ufjf.br}

\keywords{QCD, Gauge Symmetry, Standard Model}

\abstract{The interactions between gluons are important in theories such as quantum chromodynamics.   Therefore, to rediscover new features of well known methods in order to investigate the $SU(3)$ gauge group can be a new way to deal with Yang-Mills theories.  In this work we analyzed YM theories through the well known Faddeev-Jackiw formalism for constrained systems.  Besides, we showed precisely that having $U(1)$ Maxwell electromagnetic theory as a starting point we can construct $SU(3)$-like and $SU(3) \otimes SU(2) \otimes U(1)$ non-Abelian theories.}

\def\[{\left\lbrack}
\def\]{\right\rbrack}

\def\({\left(}
\def\){\right)}
\def\ni{\noindent}
\newcommand{\be}{\begin{equation}}
\newcommand{\ee}{\end{equation}}
\newcommand{\ea}{\end{eqnarray}}
\newcommand{\ba}{\begin{eqnarray}}

\begin{document}

\maketitle
\flushbottom

\pagestyle{myheadings}
\markright{QCD gauge symmetries through Faddeev-Jackiw symplectic method}

\section{Introduction}

Gauge invariance is one of the most well established concepts in theoretical physics and it is one of the main ingredients in Standard Model theory.  However, we can ask if it could have an alternative origin connected to another theory or principle.  With this motivation in mind we will show in this paper that gauge invariance could be considered an emergent concept having its origin in the algebraic formalism of a well known method that deals with constrained systems, namely, the Faddeev-Jackiw (FJ) technique.  Of course the gauge invariance idea is older than FJ's, but the results obtained here will show that the connection between both will prove that $SU(3)$ and $SU(3) \otimes SU(2) \otimes U(1)$ gauge groups, which are fundamental to important theories like QCD and Standard Model, can be obtained through FJ formalism.

To explain the FJ method, we can say that when we have to deal with constrained systems we find consistency problems that spoil the Poisson brackets algebra and consequently any quantization technique.  One of the seminal papers that solve this kind of problem was carried out by Dirac in \cite{dirac} where a consistent method for quantization was introduced for any system.  Some years later, Faddeev and Jackiw (FJ) \cite{fj} analyzed the constrained systems via symplectic approach applied on first-order Lagrangians.  After this last work, the so-called FJ symplectic approach was used and adapted to several kind of purposes.

In few words, the FJ method is an approach that is geometrically motivated.  It is based on the symplectic structure of the phase space.  The first-order characteristic allows us to obtain the Hamiltonian equations of motion from a variational principle.  Its geometric structure of the Hamiltonian phase-space will be carried out directly from the equations of motion via the inverse of the so-called symplectic two-form, if the inverse exists.
Few years after its publication, the FJ formalism was extended \cite{bw} and through the years it has been applied to different systems \cite{Oliveira}.

In this paper we will explore this alternative application of the FJ formalism in order to investigate non-Abelian gauge symmetries such as $SU(3)$ and 
$SU(3) \otimes SU(2) \otimes U(1)$.  Although these both results are very well known in the literature, our focus here is to show that the FJ technique has a wider approach than was previously aimed by its authors.  Concerning QCD, we believe that we can also analyze confinement using the symplectic formalism.  But this issue is out of the scope of this paper.

This paper is organized in such a way that in section II, using the FJ method, we developed $SU(3)$ gauge symmetries having $U(1)$ symmetry as the starting point.  In section III, beginning with $U(1)$ symmetry we have obtained $SU(3) \otimes SU(2) \otimes U(1)$ gauge symmetry.  The conclusions are described in the last section.  For pedagogical reasons, we have constructed an Appendix with the main features of QCD-like gauge theories.

\section{Constructing the $SU(3)$ non -Abelian theory}

In order to construct the $SU(3)$ non-Abelian theory let us consider as a starting point the $U(1)$ Maxwell Electromagnetic theory in four dimensions, which is described by the Lagrangian density 

\be
\label{1}
\mathcal{L} = - \frac{1}{4} {F_{\mu \nu }}{F^{\mu \nu }},
\ee

\noindent where ${F_{\mu \nu }} = {\partial _{\mu }}{A_{\nu }} - {\partial _{\nu }}{A_{\mu }}$ and the space-time metric is $g_{00} = 1$ and $g_{ii} = -1$, with i=1,2,3.

To carry out the first step of the formalism we change the original field as

\be
\label{2}
{A_{\mu }} \rightarrow {A_{\mu }^{a}},
\ee

\noindent where $a$ denotes an index belonging to $SU(3)$ symmetry group. Since we want to include a group index, we rewrite the original Lagrangian density as

\be
\label{3}
\mathcal{L} = - \frac{1}{4} {H^{a}_{\mu \nu }}{H^{a \mu \nu }},
\ee

\noindent where ${H_{\mu \nu }^{a}} = {F_{\mu \nu }^{a}} + h {{\tilde {F}}_{\mu \nu }^{a}}$, $h$ is a parameter and ${\tilde {F}}_{\mu \nu }^{a} = {\tilde {F}}_{\mu \nu }^{a} ( {A_{\alpha }^{a}} )$ is an arbitrary antisymmetric tensor. 
It is well known that ${\tau}^a$ are the generators of $SU(3)$ that obey a Lie algebra given by $[\tau^a, \tau^b] = i f^{abc}\tau^c$, where $f^{abc}$ are the structure constants of $SU(3)$ and, by introducing  the field $A_\mu = A_\mu^a (x)\tau^a$, we should have that $[A_\mu, A_\nu] = if^{abc}A_\mu^a A_\nu^b \tau^c$.

The second step of the method is to rewrite the Lagrangian density in Eq. (\ref{3}) in its first-order form. Then, using the canonical momenta ${\pi^{a}_{\mu }} = \delta \mathcal{L} / \delta {\dot {A_{\mu }^{a}}} $, we have that

\be
\label{4}
\mathcal{L} = - {\partial _{i}} {\pi _{i}^{a}} {A_{0}^{a}} - {\frac{1}{2}} {\pi _{i}^{a}} {\pi _{i}^{a}} - {\pi _{i}^{a}} {{\dot {A}}_{i}^{a}} - h {\pi _{i}^{a}} {{\tilde {F}}_{0i}^{a}}- \frac{1}{4} \left( {F^{a}_{i j}} {F^{a i j}} + 2h {F^{a}_{i j}} {{\tilde {F}}^{a i j}} + {h^{2}} {{\tilde {F}}^{a \mu \nu }} {{\tilde {F}}^{a}_{\mu \nu }} \right).
\ee

\noindent The symplectic variables are ${{\xi ^{\alpha }}} = \left( {A_{i}^{a}} \ \mbox{,} \ {\pi _{i}^{a}} \ \mbox{,} \ {A_{0}^{a}} \right)$.   The two-form matrix is singular and its zero-mode leads to the Gauss law constraint ${\Omega^{a}} = {D^{abi}} {\pi _{i}^{b}}$, where $${D^{abi}} = {\delta ^{ab}} {\partial ^{i}} + h \delta {{\tilde {F}}^{b 0i}}/ \delta {A_{0}^{a}}\,\,.$$

Substituting this constraint into the Lagrangian density, we have a new Lagrangian

\be
\label{5}
{\mathcal{L}_{new}} = {\pi _{i}^{a}} {\dot{A_{i}^{a}}} - \frac{1}{2} {\pi _{i}^{a}} {\pi _{i}^{a}} + {\Omega ^{a}} {\dot {\beta }^{a}} - \frac{1}{4} \left( {F^{a}_{i j}} {F^{a i j}} + 2h {F^{a}_{i j}} {{\tilde {F}}^{a i j}} + {h^{2}} {{\tilde {F}}^{a \mu \nu }} {{\tilde {F}}^{a}_{\mu \nu }} \right).
\ee

\noindent In this step the new symplectic variables are ${\xi ^{\alpha }_{new}} = \left( {A_{i}^{a}} \ \mbox{,} \ {\pi _{i}^{a}} \ \mbox{,} \ {\beta ^{a}} \right)$ where $\beta ^{a}$ is a Lagrange multiplier and the new two-form matrix is

\be
\label{6}
f = \left( \begin{array}{ccc} \frac{\delta {\pi _{j}^{b}}(y)}{\delta {A_{i}^{a}}(x)} - \frac{\delta {\pi _{i}^{a}}(x)}{\delta {A_{j}^{b}}(y)} & - {\delta _{ij}} {\delta ^{ab}} {\delta (x - y)} & \frac{\delta {\Omega ^{b}} (y)}{\delta {A_{i}^{a}} (x)} \\ {\delta _{ij}} {\delta ^{ab}} {\delta (x - y)} & 0 & \left( {\delta ^{ab}} {\partial _{i}^{y}} + h \frac{\delta {\tilde {F}^{a}_{0i}} (y)}{\delta {A_{0}^{b}} (y)} \right) {\delta (x-y)} \\ - \frac{\delta {\Omega ^{a}} (x)}{\delta {A_{i}^{b}} (y)} & -\left( {\delta ^{ba}} {\partial _{j}^{x}} + h \frac{\delta {\tilde {F}^{b}_{0j}} (x)}{\delta {A_{0}^{a}} (x)} \right) {\delta (x-y)} & 0 \end{array} \right).
\ee

\noindent Since we want to construct the SU(3) non-Abelian field theory, this matrix must be singular and then we perform a convenient choice for the zero-mode, namely

\be
\label{7}
{\nu ^{\alpha }} = \left( {\delta ^{ad}} {\partial _{i}^{x}} - h {f^{adc}} {A_{i}^{c}} (x) \ \mbox{,} \ - \frac{\delta {\Omega ^{d}} (x)}{\delta {A_{i}^{a}} (x)} \ \mbox{,} \ - {\delta ^{ad}} \right),
\ee

\noindent where $f^{abc}$ are the structure constants of the SU(3) group. Following the prescription of the method, the multiplication of this zero-mode by the two-form matrix in Eq. (\ref{6}) leads to

\be
\label{8}
 \frac{\delta {{\tilde {F}}^{b}_{0i}} (x)}{\delta {A_{0}^{d}} (x)} = - {f^{bdc}} {A_{i}^{c}} (x).
\ee

\noindent Then, we have
\be
\label{9} 
{\tilde {F}^{b}_{0i}} (x) = - {f^{bdc}} {A_{0}^{d}} (x){A_{i}^{c}} (x).
\ee

\noindent From this equation we conclude
\ba
\label{10} 
{H^{a}_{\mu \nu }} &=& {F^{a}_{\mu \nu }} - h{f^{abc}} {A_{\mu }^{b}} {A_{\nu }^{c}} \nonumber \\
&=& {\partial _{\mu }}{A^{a}_{\nu }} - {\partial _{\nu }}{A^{a}_{\mu }} - h{f^{abc}} {A_{\mu }^{b}} {A_{\nu }^{c}}.
\ea

\noindent which is the field strength of the $SU(3)$ theory.  We can see clearly that although we have ``put by hand" the form of the extra term in the Lagrangian in Eq. (\ref{3}) which is a valid procedure in theoretical physics, the FJ formalism fix the value of $h$ and the form of $H^{a}_{\mu\nu}$, which is the main role of the technique.  

We will see in the next section that we can use the method to even more complicated theories such as the one for the Standard Model.  As we have to introduce two structure constants, we will see that we have to add two different Lagrangians and consequently the algebraic calculations turned out to be more complicated.

\section{Constructing the $SU(3)\otimes SU(2)\otimes U(1)$ non-Abelian theory}

Now, as we intend to construct the $SU(3)\otimes SU(2)\otimes U(1)$ non-Abelian theory let us add to the Lagrangian density of the U(1) Maxwell electromagnetic theory the following terms

\be
\label{11}
{\mathcal{L}^{\, \prime}} = - \frac{1}{4} {G^{a}_{\mu \nu }}{G^{a \mu \nu }},
\ee

\be
\label{12}
{\mathcal{L}^{\, \prime \prime}} = - \frac{1}{4} {H^{b}_{\mu \nu }}{H^{b \mu \nu }},
\ee

\noindent where $G^{a}_{\mu \nu }$ and $H^{b}_{\mu \nu }$ are arbitrary tensors which will be obtained soon and $a$ and $b$ are group indexes.

Thus, following the prescription given in the last section and since Eqs. (\ref{11}) and (\ref{12}) are analogous to Maxwell Lagrangian, let us consider $G^{a}_{\mu \nu }$ and $H^{b}_{\mu \nu }$ as

\be
\label{13}
{G_{\mu \nu }^{a}} = {\partial _{\mu }} {B^{a}_{\nu }} - {\partial _{\nu }} {B^{a}_{\mu }} + g {\tilde {\chi }^{a}_{\mu \nu }},
\ee

\be
\label{14}
{H_{\mu \nu }^{b}} = {\partial _{\mu }} {C^{b}_{\nu }} - {\partial _{\nu }} {C^{b}_{\mu }} + h {\tilde {W}^{b}_{\mu \nu }},
\ee

\noindent where $g$ and $h$ are parameters, ${\tilde {\chi }^{a}_{\mu \nu }} = {\tilde {\chi }^{a}_{\mu \nu }} ({B^{a}_{\alpha }})$ and ${\tilde {W}^{b}_{\mu \nu }} = {\tilde {W}^{b}_{\mu \nu }} ({C^{b}_{\alpha }})$ are arbitrary antisymmetric tensors. Here we will also suppose that the fields ${\tilde {\chi }^{a}_{\mu \nu }}$ and ${\tilde {W}^{b}_{\mu \nu }}$ satisfy the usual non-Abelian algebra.

Taking into account our considerations, we have the following Lagrangian density
\be
\label{15}
\mathcal{L} = - \frac{1}{4} {F_{\mu \nu }}{F^{\mu \nu }} - \frac{1}{4} {G^{a}_{\mu \nu }}{G^{a \mu \nu }} - \frac{1}{4} {H^{b}_{\mu \nu }}{H^{b \mu \nu }}.
\ee

\noindent  Since we will use the FJ method, we have to write the Lagrangian in (\ref{1}) in first-order form.  It is given by
\ba
\label{16}
\mathcal{L} = &-& {\partial _{i}} {\pi _{i}} {A_{0}} - \frac{1}{2} {\pi _{i}} {\pi _{i}} - {\pi _{i}} {\dot {A}_{i}} - \frac{1}{4} {F_{ij}} {F^{ij}} - {\partial _{i}} {\pi _{i}^{a}} {B_{0}^{a}} - \frac{1}{2} {\pi _{i}^{a}} {\pi _{i}^{a}} - {\pi _{i}^{a}} {\dot {B}_{i}^{a}}  \nonumber \\
&-& g {\pi _{i}^{a}} {\tilde {\chi }_{0i}^{a}} - {\frac{1}{4}} \left( {\chi^{a}_{i j}} {\chi^{a i j}} + 2g {\chi^{a}_{i j}} {{\tilde {\chi}}^{a i j}} + {g^{2}} {{\tilde {\chi}} ^{a \mu \nu }} {{\tilde {\chi}}^{a}_{\mu \nu }} \right) \\
&-& {\partial _{i}} {\rho_{i}^{b}} {C_{0}^{b}} - \frac{1}{2} {\rho _{i}^{b}} {\rho_{i}^{b}} - {\rho_{i}^{b}} {\dot {C}_{i}^{b}} - h {\rho _{i}^{b}} {\tilde {W}^{b}_{0i}} - \frac{1}{4} \left( {W^{b}_{i j}} {W^{b i j}} + 2h {W^{b}_{i j}} {{\tilde {W}}^{b i j}} + {h^{2}} {{\tilde {W}}^{b \mu \nu }} {{\tilde {W}}^{b}_{\mu \nu }} \right)\,\,, \nonumber
\ea

\noindent where the canonical momenta are given by
\ba
\label{17}
{\pi_{0}} &=& 0, \nonumber \\
{\pi_{i}} &=& - {\partial _{0}} {A_{i}} + {\partial _{i}} {A_{0}}, \nonumber \\
{\pi _{0}^{a}} &=& 0, \nonumber \\
{\pi_{i}^{a}} &=& - {\partial_{0}} {B_{i}^{a}} + {\partial _{i}} {B_{0}^{a}} - g {\tilde {\chi }^{a}_{0 i}}, \nonumber \\
{\rho_{0}^{b}} &=& 0, \nonumber \\
{\rho _{i}^{b}} &=& - {\partial _{0}} {C_{i}^{b}} + {\partial _{i}} {C_{0}^{a}} - h {\tilde {W}^{b}_{0 i}}\,\,.
\ea

\noindent The symplectic variables are ${\xi ^{\alpha }} = \left( {A_{i}} \ \mbox{,} \ {\pi _{i}} \ \mbox{,} \ {A_{0}} \ \mbox{,} \ {B_{i}^{a}} \ \mbox{,} \ {\pi _{i}^{a}} \ \mbox{,} \ {B_{0}^{a}} \ \mbox{,} \ {C_{i}^{b}} \ \mbox{,} \ {\rho _{i}^{b}} \ \mbox{,} \ {C_{0}^{b}} \right)$ and the two-form matrix can be written as
\be
\label{18}
{f^{(0)}} = \left( \begin{array}{ccc} {\mathcal{A}^{(0)}} & 0 & 0 \\ 0 & {\mathcal{B}^{(0)}} & 0 \\ 0 & 0 & {\mathcal{C}^{(0)}} \end{array} \right),
\ee

\noindent where
\be
\label{19}
{\mathcal{A}^{(0)}} = \left( \begin{array}{ccc} 0 & - {\delta _{ij}} {\delta (x - y)} & 0 \\ {\delta _{ij}} {\delta (x - y)} & 0 & 0 \\  0 & 0 & 0 \end{array} \right),
\ee

\be
\label{20}
{\mathcal{B}^{(0)}} = \left( \begin{array}{ccc} - \frac{\delta {\pi _{i}^{c}}(x)}{\delta {B_{j}^{d}}(y)} + \frac{\delta {\pi _{j}^{d}}(y)}{\delta {B_{i}^{c}}(x)} & - {\delta _{ij}} {\delta ^{cd}} {\delta (x - y)} & 0 \\{\delta _{ij}} {\delta ^{cd}} {\delta (x - y)} & 0 & 0 \\  0 & 0 & 0 \end{array} \right),
\ee

\be
\label{21}
{\mathcal{C}^{(0)}} = \left( \begin{array}{ccc} - \frac{\delta {\rho _{i}^{c}}(x)}{\delta {C_{j}^{d}}(y)} + \frac{\delta {\rho _{j}^{d}}(y)}{\delta {C_{i}^{c}}(x)} & - {\delta _{ij}} {\delta ^{cd}} {\delta (x - y)} & 0 \\{\delta _{ij}} {\delta ^{cd}} {\delta (x - y)} & 0 & 0 \\  0 & 0 & 0 \end{array} \right).
\ee

\noindent The matrix $f^{(0)}$ in Eq. (\ref{18}) has the following zero-modes
\be
\label{2a22}
{v_{1}^{(0)}} = \left( 0 \ \mbox{,} \ 0 \ \mbox{,} \ 1 \ \mbox{,} \ 0 \ \mbox{,} \ 0 \ \mbox{,} \ 0 \ \mbox{,} \ 0 \ \mbox{,} \ 0 \ \mbox{,} \ 0 \ \right),
\ee

\be
\label{2a23}
{v_{2}^{(0)}} = \left( 0 \ \mbox{,} \ 0 \ \mbox{,} \ 0 \ \mbox{,} \ 0 \ \mbox{,} \ 0 \ \mbox{,} \ 1 \ \mbox{,} \ 0 \ \mbox{,} \ 0 \ \mbox{,} \ 0 \ \right),
\ee

\be
\label{2a24}
{v_{3}^{(0)}} = \left( \ 0 \ \mbox{,} \ 0 \ \mbox{,} \ 0 \ \mbox{,} \ 0 \ \mbox{,} \ 0 \ \mbox{,} \ 0 \ \mbox{,} \ 0 \ \mbox{,} \ 0 \ \mbox{,} \ 1 \ \right),
\ee

\noindent which lead to the constraints
\ba
\label{25}
{\omega } &=& {\partial ^{i}} {\pi _{i}} , \nonumber \\
{\Omega ^{a}} &=& {d^{aci}} {\pi _{i}^{c}} , \nonumber \\
{\Sigma ^{b}} &=& {D^{bci}} {\rho _{i}^{c}} ,
\ea

\noindent where $${d^{aci}} = {\delta ^{ac}} {\partial ^{i}} + g \frac{\delta {\tilde {\chi }^{c0i}}}{ \delta {B_{0}^{a}}}$$ and $${D^{bci}} = {\delta ^{bc}} {\partial ^{i}} + h \frac{\delta {\tilde {W}^{c0i}}}{ \delta {C_{0}^{b}}}\,\,.$$

Substituting these constraints into the Lagrangian density in Eq. (\ref{16}), we have that
\ba
 \label{26}
\mathcal{L} =  &-& \frac{1}{2} {\pi _{i}} {\pi _{i}} - {\pi _{i}} {\dot {A}_{i}} - \frac{1}{4} {F_{ij}} {F^{ij}}  - \frac{1}{2} {\pi_{i}^{a}} {\pi _{i}^{a}} - {\pi_{i}^{a}} {\dot {B}_{i}^{a}} - \frac{1}{2} {\rho_{i}^{b}} {\rho _{i}^{b}} - {\rho_{i}^{b}} {\dot {C}_{i}^{b}} + \omega {\dot {\beta }} + {\Omega ^{a}} {\dot {\beta ^{a}}} + {{\Sigma ^{b}}{\gamma ^{b}}} \nonumber \\  
&-& \frac{1}{4} \left( {\chi^{a}_{i j}} {\chi^{a i j}} + 2g {\chi^{a}_{i j}} {{\tilde {\chi }}^{a i j}} + {g^{2}} {{\tilde {\chi }}^{a \mu \nu }} {{\tilde {\chi }}^{a}_{\mu \nu }} \right) - \frac{1}{4} \left( {W_{i j}^{a}} {W^{i j}_{a}} + 2g {W_{i j}^{a}} {{\tilde {W}}^{i j}_{a}} + {g^{2}} {{\tilde {W}}^{\mu \nu }_{a}} {{\tilde {W}}_{\mu \nu }^{a}} \right) \,\,,  \nonumber \\
\ea

\noindent where $\beta $, $\beta ^{a}$ and $\gamma ^{b}$ are the Lagrange multipliers. The new symplectic variables are ${\xi ^{\alpha }} = \left( {A_{i}} \ \mbox{,} \ {\pi _{i}} \ \mbox{,} \ \beta  \ \mbox{,} \ {B_{i}^{a}} \ \mbox{,} \ {\pi _{i}^{a}} \ \mbox{,} \ {\beta ^{a}} \ \mbox{,} \ {C_{i}^{b}} \ \mbox{,} \ {\rho _{i}^{b}} \ \mbox{,} \ {\gamma ^{b}} \ \right)$ and the new two-form matrix can be written as
\be
\label{27}
{f^{(1)}} = \left( \begin{array}{ccc} {\mathcal{A}^{(1)}} & 0 & 0 \\ 0 & {\mathcal{B}^{(1)}} & 0 \\ 0 & 0 & {\mathcal{C}^{(1)}}\end{array} \right),
\ee

\noindent where
\ba
\label{28}
{\mathcal{A}^{(1)}} = \left( \begin{array}{ccc} 0 & -{\delta _{ij}} \delta (x-y) & 0 \\ {\delta _{ij}} \delta (x-y) & 0 & -{\partial _{i}^{y}} \delta (x-y) \\ 0 & {\partial _{j}^{x}} \delta (x-y) & 0  \end{array} \right),
\ea

\ba
\label{29}
{\mathcal{B}^{(1)}} = \left( \begin{array}{ccc} \delta (x-y)\frac{\delta {\pi _{j}^{d}}(y)}{\delta {B_{i}^{c}}(x)} - \frac{\delta {\pi _{i}^{c}}(x)}{\delta {B_{j}^{d}}(y)} & - {\delta _{ij}} {\delta ^{cd}} {\delta (x - y)} & \frac{\delta {\Omega ^{d}} (y)}{\delta {B_{i}^{c}} (x)} \\ {\delta _{ij}} {\delta ^{cd}} {\delta (x - y)} & 0 & -\left( {\delta ^{cd}} {\partial _{i}^{y}} + g \frac{\delta {\tilde {\chi }_{0i}^{c}} (y)}{\delta {B_{0}^{d}} (y)} \right) {\delta (x-y)} \\ - \frac{\delta {\Omega ^{c}} (x)}{\delta {B_{i}^{d}} (y)} & \left( {\delta ^{dc}} {\partial _{i}^{x}} + g \frac{\delta {\tilde {\chi }_{0i}^{d}} (x)}{\delta {B_{0}^{c}} (x)} \right) {\delta (x-y)} & 0 \end{array} \right) , \nonumber \\
\ea

\ba
\label{30}
{\mathcal{C}^{(1)}} = \left( \begin{array}{ccc} \delta (x-y)\frac{\delta {\rho _{j}^{d}}(y)}{\delta {C_{i}^{c}}(x)} - \frac{\delta {\rho _{i}^{c}}(x)}{\delta {C_{j}^{d}}(y)} & - {\delta _{ij}} {\delta ^{cd}} {\delta (x - y)} & \frac{\delta {\Sigma ^{d}} (y)}{\delta {C_{i}^{c}} (x)} \\ {\delta _{ij}} {\delta ^{cd}} {\delta (x - y)} & 0 & -\left( {\delta ^{cd}} {\partial _{i}^{y}} + h \frac{\delta {\tilde {W}_{0i}^{c}} (y)}{\delta {C_{0}^{d}} (y)} \right) {\delta (x-y)} \\ - \frac{\delta {\Sigma ^{c}} (x)}{\delta {C_{i}^{d}} (y)} & \left( {\delta ^{dc}} {\partial _{i}^{x}} + h \frac{\delta {\tilde {W}_{0i}^{d}} (x)}{\delta {C_{0}^{c}} (x)} \right) {\delta (x-y)} & 0 \end{array} \right) . \nonumber \\
\ea

Since we want to construct a gauge theory based on the $SU(3)\otimes SU(2)\otimes U(1)$ group, the matrix in Eq.(\ref{27}) must be singular and we choose three convenient zero-modes, namely
\be
\label{31}
{v_{1}^{(1)}} = \left( -{\partial _{i}} \ \mbox{,} \ 0 \ \mbox{,} \ 1 \ \mbox{,} \ 0 \ \mbox{,} \ 0 \ \mbox{,} \ 0 \ \mbox{,} \ 0 \ \mbox{,} \ 0 \ \mbox{,} \ 0 \right) \,\,,
\ee

\be
\label{32}
{v_{2}^{(1)}} = \left( 0 \ \mbox{,} \ 0 \ \mbox{,} \ 0 \ \mbox{,} \ {\delta ^{ce}} {\partial _{i}^{x}} - g {\epsilon ^{cef}} {B_{i}^{f}} (x) \ \mbox{,} \ - \frac{\delta {\Omega ^{e}} (x)}{\delta {B_{i}^{c}} (x)} \ \mbox{,} \ - {\delta ^{ce}} \ \mbox{,} \ 0 \ \mbox{,} \ 0 \ \mbox{,} \ 0 \right)\,\,,
\ee

\be
\label{33}
{v_{3}^{(1)}} = \left( 0 \ \mbox{,} \ 0 \ \mbox{,} \ 0  \ \mbox{,} \ 0 \ \mbox{,} \ 0 \ \mbox{,} \ 0 \ \mbox{,} \ {\delta ^{ce}} {\partial _{i}^{x}} - h {f^{cef}} {C_{i}^{f}} (x) \ \mbox{,} \ - \frac{\delta {\Sigma ^{e}} (x)}{\delta {C_{i}^{c}} (x)} \ \mbox{,} \ - {\delta ^{ce}} \right)\,\,,
\ee

\noindent where $\epsilon^{cef}$ are he structure constants for $SU(2)$ and $f^{cef}$ are the analogous for $U(3)$.

The first zero-mode $v_{1}^{(1)}$ does not generate new constraints and the same will occur with $v_{2}^{(1)}$ and $v_{3}^{(1)}$ since
\ba
\label{34}
\frac{\delta {\tilde {\chi }_{0i}^{c}} (x)}{\delta {B_{0}^{d}} (x)} &=& - {\epsilon ^{dce}} {B_{i}^{e}} (x) \,\,,\nonumber \\
\frac{\delta {\tilde {W}_{0i}^{c}} (x)}{\delta {C_{0}^{d}} (x)} &=& - {f^{dce}} {C_{i}^{e}} (x) \,\,.
\ea

After a straightforward calculation, we have
\ba
\label{35} 
{\tilde {\chi }_{0i}^{a}} &=& - {\epsilon ^{acd}} {B_{0}^{c}} {B_{i}^{d}} \,\,,\nonumber \\
{\tilde {W}_{0i}^{b}} &=& - {f^{bcd}} {C_{0}^{c}}{C_{i}^{d}}.
\ea

Notice that the theory must remain covariant. Therefore, we conclude that the tensors ${G^{a}_{\mu \nu }}$ and ${H^{b}_{\mu \nu }}$ have the general forms
\ba
\label{36} 
G_{\mu \nu }^{a} = {\partial _{\mu }} {B_{\nu }^{a}} - {\partial _{\nu }} {B_{\mu }^{a}} - {\epsilon ^{acd}} {B_{\mu}^{c}} (x){B_{\nu}^{d}}\,\,, \nonumber \\
H_{\mu \nu }^{b} = {\partial _{\mu }} {C_{\nu }^{b}} - {\partial _{\nu }} {C_{\mu }^{}} - {f^{bcd}} {C_{\mu}^{c}} (x){C_{\nu}^{d}} .
\ea

Hence, following the FJ methodology, the zero-modes $v_{1}^{(1)}$, $v_{2}^{(1)}$ and $v_{3}^{(1)}$ are the generators of infinitesimal gauge transformations, given by
\ba
\label{37} 
\delta {A_{\mu }} &=& - {\partial _{\mu }} \varepsilon\,\,, \nonumber \\
\delta {B^{a}_{\mu }} &=& - {\partial _{\mu }} {\varepsilon^{a}} - g {\epsilon ^{acd}} {B^{c}_{\mu }} {\varepsilon^{d}}\,\,, \nonumber \\
\delta {C^{b}_{\mu }} &=& - {\partial _{\mu }} {\varepsilon^{b}} - h {f^{bcd}} {C^{b}_{\mu }} {\varepsilon^{d}}\,\,.
\ea

Observe that we have used conveniently the arbitrary zero-mode property.  The main objective here was to introduce the structure constants of $SU(2)$ and $SU(3)$ gauge groups.  However, we have to say that there is one constraint concerning the zero-mode.  It is the constraint of physical reality.  Namely,  one have to observe if the chosen zero-mode did not introduce unphysical ingredients in the theory like tachyons. To be sure, one have to make an analysis of the spectrum of the final theory.  In our case, the final theories are very well known and to perform such an analysis is unnecessary.

We would like to comment that both results of the last sections can lead us to conjecture that, since the algebra of the FJ method could provide both $SU(3)$ and $SU(3) \otimes SU(2) \otimes U(1)$ theory together with its gauge symmetries, we imagine if the FJ procedure could be considered an alternative origin of the gauge symmetry concept itself.  When we deal with constrained systems, of course.  But then, to extend the gauge symmetry idea (obtained with FJ technique) to unconstrained models would be straightforward.  One possible next step would be also to investigate if there is a connection between the FJ approach and Noether theorem.

\section{Conclusions}

The FJ symplectic formalism was constructed at the final part of the eighties and had a great development in the nineties.  From that point until now we have observed the growing of its applications and perspectives.   And the construction of an extension of its main structure in the literature.  To sum up the FJ method, the transformation of a higher-order Lagrangian into first-order is done by augmenting the configuration space with auxiliary variables.  After that we have a new set of variables, the symplectic variables.  One of the strongest points of the method is the relative arbitrariness of the zero-mode.  This permits one to construct a whole family of gauge symmetries which is very useful.  However, as we comment above, we have to analyze the spectrum of the final theory in order to avoid unphysical objects being introduced in the process.

In this work we took advantage of this arbitrariness property.  Firstly, in section II we have obtained the $SU(3)$ gauge theory using only the FJ method.  In the following section after constructing the symplectic matrix we could construct the zero-mode that have provided both $SU(2)$ and $SU(3)$ structure constants.  In this way our final theory has  $SU(3)\otimes SU(2) \otimes U(1)$ gauge symmetry and we present the respective gauge transformations.

Since the gauge symmetry is an underlying component of the Standard Model, which has an $SU(3)\otimes SU(2)\otimes U(1)$ symmetry we can imagine, as a next step, what the FJ method can say about the Higgs mechanism.  It is an ongoing research.

\section{Appendix - QCD: the $SU(3)$ approach}

Our motivation in this section is, besides to provide a brief and pedagogical presentation of the main parts of QCD gauge theory, to carry out some comments about the gluon gauge field that helps the physical understanding of the mathematics developed in section III.

One of the most beautiful concepts in theoretical physics is the interactions are governed by symmetry principles.  It is well known that the structure of QCD can be originated form local gauge invariance.  The symmetry group is the $SU(3)$ group of phase transformation on the quark color fields.

Let us write the free Lagrangian given by
\be
\label{ApA}
{\cal L}_0\,=\,\bar{q}_j\,(i\gamma^{\mu} \partial_{\mu}\,-\,m ) q_j
\ee

\ni where $q_1\,,\,q_2$ and $q_3$ indicate the three color fields.

To guarantee that ${\cal L}_0$ is invariant under local phase transformations given by
\be
\label{ApB}
q(x) \rightarrow U q(x) \equiv e^{i\alpha_a (x) T_a}\,q(x)
\ee

\ni where $U$ is an arbitrary $3 \times 3$ unitary matrix.  In Eq. (\ref{ApB}) we assumed a summation over the repeated $a$.  We define $T_a\:(a=1, \ldots,8 )$ as a set of linearly independent traceless $3 \times 3$ matrices.  Concerning $\alpha_a$, they are group parameters.  The algebra constructed with the generators $T_a$ is non-Abelian given by $\[T_a , T_b \]\,=\,if_{abc}\,T_c$, where $f_{abc}$ are the so-called the structure constants of the group.

The problem here is to make the Lagrangian in (\ref{ApA}) invariant under $SU(3)$ local gauge transformations
\ba
&& q(x) \rightarrow \Big[ 1\,+\,i\alpha_a(x) T_a \Big]\,q(x) \\ \label{ApC}
&& \partial_{\mu} \rightarrow \Big( 1\,+\,i\alpha_a T_a \Big)\partial_{\mu} \label{ApD}
\ea

\ni where the last term in (\ref{ApD}) spoils the invariance of ${\cal L}_0$.  The construction of an invariant Lagrangian requires the introduction of (eight) gauge fields $G^a_{\mu}$ each transforming as
\be
\label{ApE}
G^a_{\mu} \rightarrow g^a_{\mu}\,-\, \frac 1g \partial_{\mu} \alpha_a
\ee

\ni and a covariant derivative
\be
\label{ApF}
D_{\mu}\,=\,\partial_{\mu}\,+\,i g T_a G^a_{\mu}
\ee

\ni in order to substitute $\partial_{\mu} \rightarrow D_{\mu}$ in Lagrangian (\ref{ApA}) and obtain
\be
\label{ApG}
{\cal L}\,=\, \bar{q}\, \Big( i\gamma^{\mu} \partial_{\mu}\,-\,m \Big) q\,-\, g \Big( \bar{q} \gamma^{\mu} T_a q \Big) G^a_{\mu}\,\,.
\ee

However, due to the non-Abelian nature of the gauge transformation, it is not simple to produce a gauge invariant Lagrangian.  Hence, to obtain the gauge invariance, we have to rewrite (\ref{ApE}) such that
\be
\label{ApH}
G^a_{\mu}  \rightarrow G^a_{\mu} \,-\,\frac 1g \partial_{\mu} \alpha_a \,-\,f_{abc} \alpha_b G^c_{\mu}
\ee

\ni and the consequence is to add a kinetic energy term for each of the $G^a_{\mu}$ fields.  So, the resulting gauge invariant QCD Lagrangian is
\be
\label{ApI}
{\cal L}\,=\, \bar{q}\, \Big( i\gamma^{\mu} \partial_{\mu}\,-\,m \Big) q\,-\, g \Big( \bar{q} \gamma^{\mu} T_a q \Big) G^a_{\mu}\,-\,\frac 14 G^a_{\mu\nu}\,G^{\mu\nu}_a\,\,,
\ee

\ni where $G^a_{\mu}$ is given by
\be
\label{ApJ}
G^a_{\mu\nu}\,=\,\partial_{\mu} G^a_{\nu}\,-\,\partial_{\nu}G^a_{\mu}\,-\,g f_{abc} G^b_{\mu} G^c_{\nu}\,\,.
\ee

\ni From Eq. (\ref{ApI}) we can see that, analogously to the photon, the gluon field $G_{\mu}^a$ is required, by gauge invariance, to be massless.  The Lagrangian in Eq. (\ref{ApI}) is for interacting colored quarks $q$ and vector gluons $G_{\mu}^a$, with the coupling being specified by $g$.  Or, in other words, it describes three Dirac fields with equal mass (the three colors of a given quark flavor) interacting with eight massless vector fields, namely, the gluons.  The Dirac fields constitute eight color currents which have the role of being the sources for the color fields, $A_{\mu}^a$.  As the phase transformations of the three quark color fields can vary, we have that eight vector gluon fields are needed to compensate all possible changes, remembering that $G^a_{\mu}$ has $a=1, \ldots,8$.

Another feature can be seen in the field strength tensor $G^a_{\mu\nu}$ written in Eq. (\ref{ApJ}).  The kinetic energy term in Eq. (\ref{ApI}) is not purely kinetic but includes an induced self-interaction between gauge bosons.  This is a manifestation of the non-Abelian property of the gauge group.  The gauge invariance establish, in a unique way, the structure of the gluon self-coupling terms.  But notice that there is only one coupling $g$.

\acknowledgments  EMCA and WO would like to thank Conselho Nacional de Desenvolvimento Cient\' ifico e Tecnol\'ogico (CNPq) for partial financial support.

\end{document}